\documentclass[aps,
prd,
reprint,
preprintnumbers,
superscriptaddress,
floatfix]{revtex4-1}

\usepackage{graphicx}
\usepackage{xspace}
\usepackage{units}
\usepackage{hyperref}
\usepackage{lineno}
\usepackage{verbatim}

\newcommand{\figurescale}{0.99\columnwidth}

\def\nova{NOvA\xspace}

\begin{document}

\preprint{FERMILAB-PUB-20-472-ND}

\title{Search for Slow Magnetic Monopoles with the NOvA Detector on the Surface}

\newcommand{\ANL}{Argonne National Laboratory, Argonne, Illinois 60439, 
USA}
\newcommand{\ICS}{Institute of Computer Science, The Czech 
Academy of Sciences, 
182 07 Prague, Czech Republic}
\newcommand{\IOP}{Institute of Physics, The Czech 
Academy of Sciences, 
182 21 Prague, Czech Republic}
\newcommand{\Atlantico}{Universidad del Atlantico,
Carrera 30 No. 8-49, Puerto Colombia, Atlantico, Colombia}
\newcommand{\BHU}{Department of Physics, Institute of Science, Banaras 
Hindu University, Varanasi, 221 005, India}
\newcommand{\UCLA}{Physics and Astronomy Department, UCLA, Box 951547, Los 
Angeles, California 90095-1547, USA}
\newcommand{\Caltech}{California Institute of 
Technology, Pasadena, California 91125, USA}
\newcommand{\Cochin}{Department of Physics, Cochin University
of Science and Technology, Kochi 682 022, India}
\newcommand{\Charles}
{Charles University, Faculty of Mathematics and Physics,
 Institute of Particle and Nuclear Physics, Prague, Czech Republic}
\newcommand{\Cincinnati}{Department of Physics, University of Cincinnati, 
Cincinnati, Ohio 45221, USA}
\newcommand{\CSU}{Department of Physics, Colorado 
State University, Fort Collins, CO 80523-1875, USA}
\newcommand{\CTU}{Czech Technical University in Prague,
Brehova 7, 115 19 Prague 1, Czech Republic}
\newcommand{\Dallas}{Physics Department, University of Texas at Dallas,
800 W. Campbell Rd. Richardson, Texas 75083-0688, USA}
\newcommand{\DallasU}{University of Dallas, 1845 E 
Northgate Drive, Irving, Texas 75062 USA}
\newcommand{\Delhi}{Department of Physics and Astrophysics, University of 
Delhi, Delhi 110007, India}
\newcommand{\JINR}{Joint Institute for Nuclear Research,  
Dubna, Moscow region 141980, Russia}
\newcommand{\FNAL}{Fermi National Accelerator Laboratory, Batavia, 
Illinois 60510, USA}
\newcommand{\UFG}{Instituto de F\'{i}sica, Universidade Federal de 
Goi\'{a}s, Goi\^{a}nia, Goi\'{a}s, 74690-900, Brazil}
\newcommand{\Guwahati}{Department of Physics, IIT Guwahati, Guwahati, 781 
039, India}
\newcommand{\Harvard}{Department of Physics, Harvard University, 
Cambridge, Massachusetts 02138, USA}
\newcommand{\Houston}{Department of Physics, 
University of Houston, Houston, Texas 77204, USA}
\newcommand{\IHyderabad}{Department of Physics, IIT Hyderabad, Hyderabad, 
502 205, India}
\newcommand{\Hyderabad}{School of Physics, University of Hyderabad, 
Hyderabad, 500 046, India}
\newcommand{\IIT}{Illinois Institute of Technology,
Chicago IL 60616, USA}
\newcommand{\Indiana}{Indiana University, Bloomington, Indiana 47405, 
USA}
\newcommand{\INR}{Institute for Nuclear Research of Russia, Academy of 
Sciences 7a, 60th October Anniversary prospect, Moscow 117312, Russia}
\newcommand{\Iowa}{Department of Physics and Astronomy, Iowa State 
University, Ames, Iowa 50011, USA}
\newcommand{\Irvine}{Department of Physics and Astronomy, 
University of California at Irvine, Irvine, California 92697, USA}
\newcommand{\Jammu}{Department of Physics and Electronics, University of 
Jammu, Jammu Tawi, 180 006, Jammu and Kashmir, India}
\newcommand{\Lebedev}{Nuclear Physics and Astrophysics Division, Lebedev 
Physical 
Institute, Leninsky Prospect 53, 119991 Moscow, Russia}
\newcommand{\Magdalena}{Universidad del Magdalena, Carrera 32 No 22 –-
 08 Santa Marta, Colombia}
\newcommand{\MSU}{Department of Physics and Astronomy, Michigan State 
University, East Lansing, Michigan 48824, USA}
\newcommand{\Crookston}{Math, Science and Technology Department, University 
of Minnesota Crookston, Crookston, Minnesota 56716, USA}
\newcommand{\Duluth}{Department of Physics and Astronomy, 
University of Minnesota Duluth, Duluth, Minnesota 55812, USA}
\newcommand{\Minnesota}{School of Physics and Astronomy, University of 
Minnesota Twin Cities, Minneapolis, Minnesota 55455, USA}
\newcommand{\Mississippi}{University of Mississippi, University, Mississippi 38677, USA}
\newcommand{\NISER}{National Institute of Science Education and Research,
Khurda, 752050, Odisha, India}
\newcommand{\Oxford}{Subdepartment of Particle Physics, 
University of Oxford, Oxford OX1 3RH, United Kingdom}
\newcommand{\Panjab}{Department of Physics, Panjab University, 
Chandigarh, 160 014, India}
\newcommand{\Pitt}{Department of Physics, 
University of Pittsburgh, Pittsburgh, Pennsylvania 15260, USA}
\newcommand{\QMU}{School of Physics and Astronomy,
 Queen Mary University of London,
London E1 4NS, United Kingdom}
\newcommand{\RAL}{Rutherford Appleton Laboratory, Science 
and 
Technology Facilities Council, Didcot, OX11 0QX, United Kingdom}
\newcommand{\SAlabama}{Department of Physics, University of 
South Alabama, Mobile, Alabama 36688, USA} 
\newcommand{\Carolina}{Department of Physics and Astronomy, University of 
South Carolina, Columbia, South Carolina 29208, USA}
\newcommand{\SDakota}{South Dakota School of Mines and Technology, Rapid 
City, South Dakota 57701, USA}
\newcommand{\SMU}{Department of Physics, Southern Methodist University, 
Dallas, Texas 75275, USA}
\newcommand{\Stanford}{Department of Physics, Stanford University, 
Stanford, California 94305, USA}
\newcommand{\Sussex}{Department of Physics and Astronomy, University of 
Sussex, Falmer, Brighton BN1 9QH, United Kingdom}
\newcommand{\Syracuse}{Department of Physics, Syracuse University,
Syracuse NY 13210, USA}
\newcommand{\Tennessee}{Department of Physics and Astronomy, 
University of Tennessee, Knoxville, Tennessee 37996, USA}
\newcommand{\Texas}{Department of Physics, University of Texas at Austin, 
Austin, Texas 78712, USA}
\newcommand{\Tufts}{Department of Physics and Astronomy, Tufts University, Medford, 
Massachusetts 02155, USA}
\newcommand{\UCL}{Physics and Astronomy Department, University College 
London, 
Gower Street, London WC1E 6BT, United Kingdom}
\newcommand{\Virginia}{Department of Physics, University of Virginia, 
Charlottesville, Virginia 22904, USA}
\newcommand{\WSU}{Department of Mathematics, Statistics, and Physics,
 Wichita State University, 
Wichita, Kansas 67206, USA}
\newcommand{\WandM}{Department of Physics, William \& Mary, 
Williamsburg, Virginia 23187, USA}
\newcommand{\Wisconsin}{Department of Physics, University of 
Wisconsin-Madison, Madison, Wisconsin 53706, USA}
\newcommand{\deceased}{Deceased.}
\affiliation{\ANL}
\affiliation{\Atlantico}
\affiliation{\BHU}
\affiliation{\Caltech}
\affiliation{\Charles}
\affiliation{\Cincinnati}
\affiliation{\Cochin}
\affiliation{\CSU}
\affiliation{\CTU}
\affiliation{\DallasU}
\affiliation{\Delhi}
\affiliation{\FNAL}
\affiliation{\UFG}
\affiliation{\Guwahati}
\affiliation{\Harvard}
\affiliation{\Houston}
\affiliation{\Hyderabad}
\affiliation{\IHyderabad}
\affiliation{\IIT}
\affiliation{\Indiana}
\affiliation{\ICS}
\affiliation{\INR}
\affiliation{\IOP}
\affiliation{\Iowa}
\affiliation{\Irvine}
\affiliation{\JINR}
\affiliation{\Lebedev}
\affiliation{\Magdalena}
\affiliation{\MSU}
\affiliation{\Duluth}
\affiliation{\Minnesota}
\affiliation{\Mississippi}
\affiliation{\NISER}
\affiliation{\Panjab}
\affiliation{\Pitt}
\affiliation{\QMU}
\affiliation{\SAlabama}
\affiliation{\Carolina}
\affiliation{\SDakota}
\affiliation{\SMU}
\affiliation{\Stanford}
\affiliation{\Sussex}
\affiliation{\Syracuse}
\affiliation{\Texas}
\affiliation{\Tufts}
\affiliation{\UCL}
\affiliation{\Virginia}
\affiliation{\WSU}
\affiliation{\WandM}
\affiliation{\Wisconsin}

\author{M.~A.~Acero}
\affiliation{\Atlantico}

\author{P.~Adamson}
\affiliation{\FNAL}



\author{L.~Aliaga}
\affiliation{\FNAL}

\author{T.~Alion}
\affiliation{\Sussex}

\author{V.~Allakhverdian}
\affiliation{\JINR}




\author{N.~Anfimov}
\affiliation{\JINR}


\author{A.~Antoshkin}
\affiliation{\JINR}


\author{E.~Arrieta-Diaz}
\affiliation{\Magdalena}

\author{L.~Asquith}
\affiliation{\Sussex}


\author{A.~Aurisano}
\affiliation{\Cincinnati}


% ...

\collaboration{The NOvA Collaboration}
\noaffiliation

\begin{abstract}
We report a search for a magnetic monopole component of the cosmic-ray flux in a 95-day exposure of the NOvA experiment's Far Detector, a 14\,kt segmented liquid scintillator detector designed primarily to observe GeV-scale electron neutrinos.  No events consistent with monopoles were observed, setting an upper limit on the flux of $2\times 10^{-14}\,\mathrm{cm^{-2}s^{-1}sr^{-1}}$ at 90\% C.L. for monopole speed $6\times 10^{-4} < \beta < 5\times 10^{-3}$ and mass greater than $5\times 10^{8}$\,GeV.  Because of NOvA's small overburden of 3~meters-water equivalent, this constraint covers a previously unexplored low-mass region.
\end{abstract}

\maketitle

\nolinenumbers

\section{Introduction}
\label{sec:introduction}

Magnetically charged particles were hypothesized by Dirac in 1931~\cite{Dirac:1931kp} and are generically predicted by grand unified theories (GUTs)~\cite{Tanabashi:2018oca,tHooft:1974kcl,Polyakov:1974ek}.  Although GUT-scale monopoles of $\sim$$10^{17}$--$10^{18}$\,GeV are often assumed, recent theoretical work suggests possible masses as light as $\sim$$10^7$\,GeV~\cite{Kephart:2001ix, Tanabashi:2018oca}.  Searches over the past century for a monopole component of the cosmic ray flux have yet to find convincing evidence~\cite{Groom:1986ps}.  Slow-moving ($\beta < 0.01$) GUT-scale monopoles have been ruled out by underground experiments~\cite{macro}.  Weaker limits exist for slow monopoles in the range $10^5\,\mathrm{GeV} < m < 10^{12}$\,GeV from mountaintop experiments~\cite{slim}.  In this paper, we focus on the possibility that there is a flux of slow cosmic-ray magnetic monopoles.  As a large low-elevation surface detector, the NOvA Far Detector is sensitive to a combination of monopole masses and speeds not previously accessible.


The \nova experiment primarily measures the oscillation of muon neutrinos~\cite{Acero:2019ksn}.  The measurement of this oscillation in neutrinos and antineutrinos gives information about the mixing parameters $\theta_{23}$ and $\Delta m^{2}_{32}$, the neutrino mass hierarchy, and the CP-violating phase $\delta_{CP}$ of the PMNS matrix.  \nova uses two detectors, the 0.3\,kt Near Detector underground at Fermilab, 50\,km west of downtown Chicago, IL, and the 14\,kt Far Detector (FD) near Ash River, MN.  A beam of muon neutrinos produced at Fermilab travels through the Earth to the FD.

Due to its surface location, monopoles with $m \gtrsim 10^8$\,GeV can reach the FD without being absorbed by the atmosphere or overburden, while its size and trigger design allow identification of slow tracks.  Compared to a dedicated underground monopole detector, \nova is not optimized for monopole detection and must contend with a large cosmic muon background.

\begin{figure}
\includegraphics[width=\columnwidth]{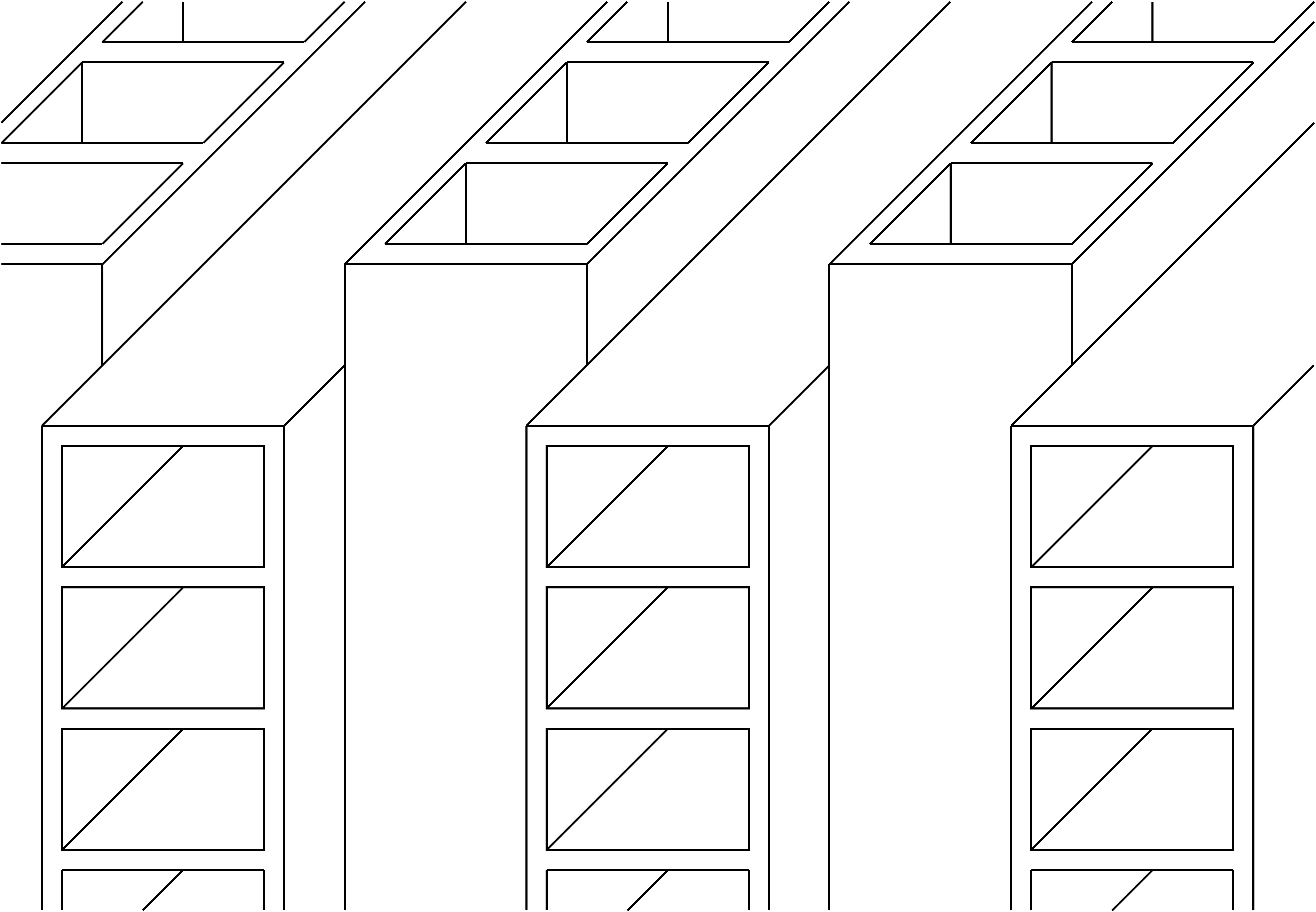}
\caption{Schematic of a corner of the NOvA detector.  The $z$ direction is to the right, perpendicular to the 15.5~meter-long cells, the ends of which are shown.}
\label{fig:det}
\end{figure}

The \nova FD has been described previously~\cite{novatdr}.  In brief, the detector is on the surface, with a concrete and barite overburden of 3~meters-water equivalent. It is a segmented detector with dimensions \unit[15.5]{m} by \unit[15.5]{m} by \unit[59.8]{m}, consisting of 896 planes of 384 plastic cells each filled with organic liquid scintillator. Each cell is \unit[15.5]{m} by \unit[4]{cm} by \unit[6]{cm}.  Planes alternate between $x$ and $y$ orientations (see Fig.~\ref{fig:det}), with signals acquired from two projected views, $xz$ and $yz$, separately.  The $x$-direction points $28^\circ$ south of west, $y$ is vertical, and $z$ is the long axis of the detector such that the three form a right-handed coordinate system.  Light produced in the cells by ionizing particles is collected by a loop of wavelength-shifting fiber and converted to electrical signals by avalanche photodiodes (APDs).  All APD signals are continuously digitized at 2\,MHz.  In each cell, samples that rise above a threshold defined to exclude the majority of noise are retained for further trigger processing.  Such a sample from one cell is called a ``hit.''  A cell can satisfy the criteria for a hit as often as once every three clock cycles, or 1.5\,$\mu$s.  Each hit provides a 2D position; 3D trajectories are reconstructed using hits from the two views.


The remainder of this paper is organized as follows. We lay out our assumptions about monopole interactions and our detector simulation, which are used to determine detection efficiency, in Section~\ref{sec:simulation}.  NOvA's dedicated monopole trigger is described in Section~\ref{sec:trigger}, the offline event selection in Section~\ref{sec:selection}, and we give our results in Section~\ref{sec:results}.

\section{Simulation}\label{sec:simulation}

We used simulation to determine the efficiency of selecting monopole tracks across the range of speeds $\beta = 10^{-4}$ to $10^{-2}$.  We did not simulate proton decay catalyzed by monopoles~\cite{Rubakov:1981rg} and in our analysis we assume it does not occur at a significant rate, leaving instead only an ionization signal.  Ahlen and Kinoshita have calculated the energy deposition for non-catalyzing slow monopoles ($\beta < 10^{-2}$)~\cite{Ahlen:1982mx}:
\begin{equation}
\frac{dE}{dx} = a N_{e}^{2/3} 
\left[
\ln \left(b N_{e}^{1/3}\right) - \frac{1}{2}
\right] \beta,
\label{eq:dedx}
\end{equation}
where $N_{e}$ is the electron density, which depends on the material.  The constants $a$ and $b$ are material-independent and are defined as:
\begin{equation}
a = \frac{2 \pi g^{2} e^{2}}{\hbar c (3 \pi^2)^{1/3}} \quad \quad \quad
b = 2 (3 \pi^2)^{1/3} a_{0},
\end{equation}
where $e$ is the fundamental electric charge, $\hbar$ is the reduced Planck constant, $c$ is the speed of light, and $a_{0}$ is the Bohr radius.  For this search, the monopole charge was assumed to be the Dirac charge, $g = e/2\alpha=\hbar c/2e$, where $\alpha$ is the fine structure constant.  The monopole was assumed to have no electric charge.  This is not a unique choice; other assumptions would lead to higher or lower predicted detection efficiencies.  For instance, a magnetic monopole that also carried an electric charge, or with a multiple of the Dirac charge, would, in general, be easier to detect.  An exotic slow particle with only electric charge, e.g. a microscopic black hole, would be detected with higher or lower efficiency depending on its charge.

If, contrary to our assumptions, monopoles were to catalyze proton decay, then this energy deposition estimate would be conservative.  Generally, catalysis would slightly increase detection efficiency by causing more hits over threshold.  However, should it occur at a very high rate, it could cause reconstructed monopole tracks to be rejected as non-linear (see Section~\ref{sec:selection}).  Catalysis would also increase the mass threshold of the search by reducing the range of monopoles in the Earth and in the atmosphere.

A monopole traversing the FD would deposit energy in the liquid scintillator, most of which is visible through ionization and atomic excitation.  \nova's scintillator is mostly composed of mineral oil (solvent) and pseudocumene (scintillant)~\cite{novascint}; its electron density is $2.9 \times 10^{23}\,\mathrm{cm}^{-3}$.  Using Eq.~\ref{eq:dedx}, this yields $dE/dx = (12\,\mathrm{GeV/cm}) \beta$, which is valid for $10^{-4} < \beta < 10^{-2}$.  Although there is substantial theoretical uncertainty on the value of $dE/dx$, we used this nominal value to set limits in this search.  At $\beta \approx 1.5\times10^{-4}$, a monopole in this model has the same $dE/dx$ as a minimum-ionizing muon.

In the absence of information about the directional distribution of monopoles, an isotropic flux was assumed.  Geant4~\cite{geant} was used to track monopoles through the detector's geometry, using the same detailed detector model that has been used to analyze data for NOvA's neutrino oscillation results.  For the mass and speed ranges we considered, the energy lost by a monopole in the detector would be negligible compared to its initial kinetic energy, and it was assumed that its speed remains constant. Since energy deposition is not a function of mass, using this assumption it is not necessary to specify a mass in the simulation.  Scintillation light production, propagation and detection is modeled using experiment-specific code~\cite{novasim}.

The \nova detectors were designed to measure energy deposition of particles with $\beta$ near 1, which traverse the width of each cell in the detector in under a nanosecond.  No consideration was given in the design phase to particles depositing a similar amount of energy over as much as a microsecond, as monopoles at the lower end of our sensitivity would.  In order to verify our simulation of the detector response to such slow signals, we performed a dedicated test stand measurement which imitated the signature of monopole signals by exposing APDs read out by NOvA electronics to light pulses generated by LEDs.  The pulses had lengths that corresponded to the cell-crossing time of monopoles of various speeds and intensities corresponding to the expected monopole $dE/dx$.  The measured signal agreed with our simulation of detector electronics.  Several sources of systematic error relating to the test stand were considered, and we assigned a total uncertainty of $\pm10\%$ to the measurement. To conservatively account for the possibility that signals from slow energy depositions were overestimated in our detector simulation, we reduced the simulated detector response in the FD by 10\%.

Each simulated monopole was combined with 5\,ms of zero bias data from the FD (i.e. data with typical running conditions, saved to permanent storage without regard to its content).  The trigger operates on data blocks of this length.  This results in an event that contains both the simulated monopole and real detector activity.  Samples at various monopole speeds were used to measure how well the search algorithm can identify slow monopoles and differentiate them from the cosmic-ray background.  Twenty thousand simulated monopoles were generated at each tenth of a decade in speed across the relevant range of $\beta$.

\section{Online Trigger Algorithm}
\label{sec:trigger}

Using NOvA's data-driven trigger system~\cite{Norman:2015ete}, the FD is able to isolate interesting physics signals among 150\,kHz of cosmic rays.  This trigger system operates entirely in software and is able to perform arbitrary analyses of incoming data, although the complexity is limited by CPU time. The event topology in this search is a straight track traversing the FD, in any direction, with a speed that is a small fraction of the speed of light.  The trigger was optimized for $\beta=10^{-3}$ magnetic monopoles.

\begin{figure}
\centerline{\includegraphics[width=\columnwidth]{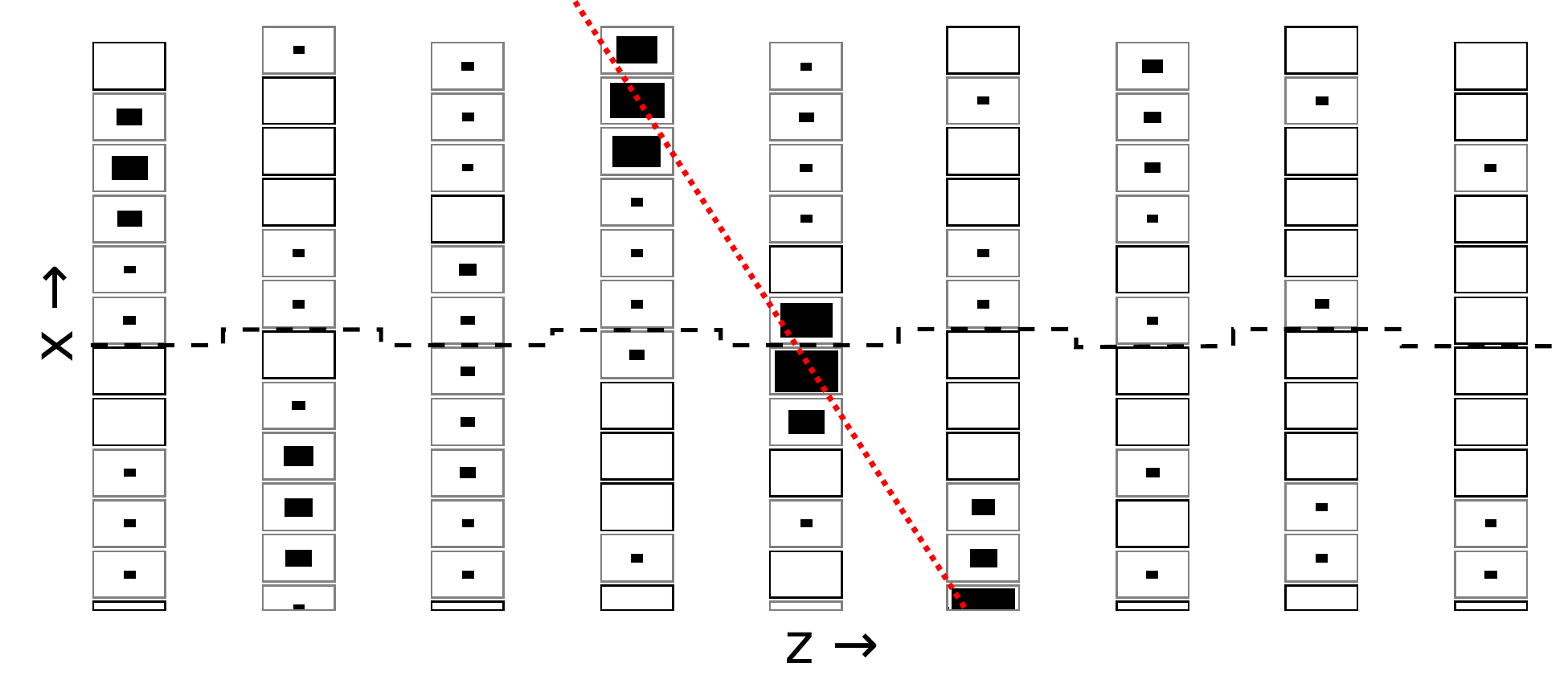}}
\caption{Hit selection in the trigger algorithm.  Cells near the $+x$ edge in the $xz$ view are shown.  Filled boxes represent hits in a 5\,ms window; size is proportional to signal strength.  The dotted red line shows the path of a simulated monopole; hits off this line are zero bias data.  Hits above the horizontal dashed line are considered to be on the detector edge.}
\label{fig:trigger_schematic}
\end{figure}

Pairs of hits within 2\,$\mu$s of each other and in neighboring $xz$ and $yz$ planes are grouped together.  These define 3D positions.  Each 3D pair that defines a position within six cells or five planes of the surface of the detector (see Fig.~\ref{fig:trigger_schematic}) is retained for further processing.  Using the $xz$ view alone for CPU efficiency, the trigger forms track seeds consisting of two selected hits on different detector faces and a time difference consistent with originating from a particle with $10^{-4.4} < \beta_{2D} < 10^{-2.3}$ in the $xz$ view.  The lower speed limit was set to approximately correspond to when the average monopole track would no longer be contained within a 5\,ms data block; CPU time was saved by not considering slower particles.  This 2D speed can correspond to a 3D speed as large as $\beta \approx 10^{-2.0}$.

For each track seed, the algorithm identifies hits that lie on a 20 cell (80\,cm) wide ``road'' with its center along the line connecting the seed hits.  It then looks for gaps between adjacent hits on the road and identifies the maximum plane gap (in the $z$-direction) and the maximum cell gap (in the $x$-direction).  If there is a plane gap larger than 30 planes (200\,cm) or a cell gap larger than 20 cells, the algorithm rejects the track seed.  Otherwise, a time window of data containing the track seed plus 4\,$\mu$s before and afterwards is written out to permanent storage.  Because a true monopole would generally produce many track seeds that pass the algorithm's selection, only every tenth track seed is checked, to save CPU time. Since a monopole that created many track seeds would also have many hits on its road, the check for gaps rejects background without significantly affecting the signal.

\begin{figure}
    \centering
    \includegraphics[width=\figurescale]{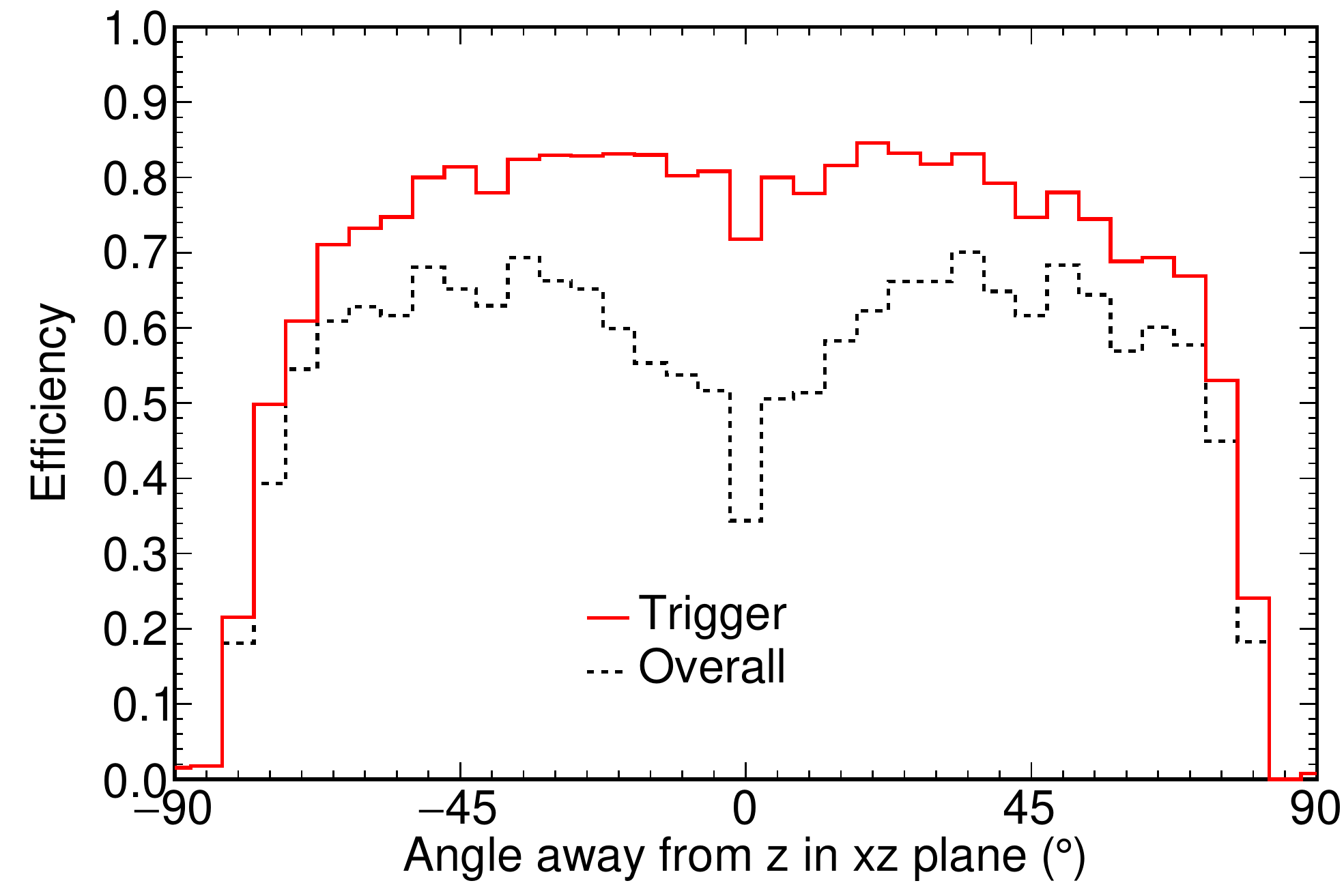}
    \caption{Trigger and overall selection efficiency as a function of angle in the $xz$ view, for monopoles that cross at least 10\,m of the detector with $\beta=10^{-3}$.  Efficiency in the $yz$ view has the same form.}
    \label{fig:thetax}
\end{figure}

Integrated over all angles, the efficiency for triggering on a monopole with $\beta = 10^{-3}$ that intersects the detector with a true crossing length of at least 10\,m is 68\% (see Fig.~\ref{fig:thetax}).  This efficiency is the result of using the conservative lower bound on the efficiency of the readout electronics discussed in Section~\ref{sec:simulation}.  If we did not use the lower bound, the estimated efficiency would be higher: 72\% at $\beta=10^{-3}$.  Most of the efficiency loss at the trigger level is caused by the need for the monopole to intersect enough cells in both $xz$ and $yz$ views for its 3D pairs on the surface to be examined.  

\section{Event Selection}
\label{sec:selection}


The initial stage of offline event selection was track reconstruction, first of speed-of-light tracks, then of slow tracks.  Speed-of-light tracks are primarily cosmic-ray muons.  Such a particle takes 50\,ns to traverse the height or width of the detector and 200\,ns to traverse the length; the hit timing resolution is typically 20\,ns.  Candidate tracks have hundreds of hits, making speed-of-light tracks easily distinguishable from the slow tracks of interest in this search. To find speed-of-light tracks, hits were clustered using their proximity in time and space. Within these clusters, straight lines consisting of several hits were identified and joined together to form tracks.  All hits belonging to such tracks were removed from consideration for monopole track reconstruction.  Since removed tracks can overlap a potential monopole track, some true monopole hits may be discarded at this step, reducing the search efficiency.  Our simulations showed fewer than 1\% of monopole hits would be discarded in this way across the range of $\beta$ considered.

Hits were then removed if they are separated from all other hits by at least two planes and two cells in their respective views.  This removal ensures that sparse tracks are not reconstructed out of stray hits arising from radioactive decays, low energy components of cosmic ray showers, hot channels, and other sources of background that are not reconstructed as speed-of-light tracks.  The remaining hits were reconstructed using the Hough tracking algorithm~\cite{hough} to identify straight line objects.  (Since the monopoles under investigation are so heavy, they do not undergo significant multiple scattering and should appear as perfectly straight lines up to the detector's resolution.) A line was fitted to each such collection of hits.  A candidate was required to meet the following three basic requirements, as well as others described below.  The track must:

\begin{enumerate}
\item Have at least 20 hits in each view;
\item \label{req:planes} Cross at least 10 planes in each view; and
\item Have a reconstructed length of at least 10\,m.
\end{enumerate}

Some efficiency for tracks with a small angles relative to $z$ is lost from requirement \ref{req:planes}, as shown in Fig.~\ref{fig:thetax}.

The speed, linear correlation coefficient, and time gap fraction (defined below) were calculated for these candidate monopole tracks.  As the trigger algorithm searches for monopoles with $\beta < 0.01$, and the analysis strategy is optimized for slow monopoles, any candidate monopole track reconstructed with a speed above this was rejected.

Since slow monopoles are not expected to be highly ionizing, distinguishing characteristics used in this search were the straightness of their tracks and their consistent slow speed. The standard linear regression correlation coefficient ($r^2$) was calculated for hits in $xt$ and $yt$ separately.  A true monopole track would have $r^2$ close to unity.   The minimum of $r^2_{xt}$ and $r^2_{yt}$ is called $r^2_\mathrm{min}$.

A potential background was reconstruction failures in which two speed-of-light cosmic rays are identified as a single monopole track.  Such a background track would have a cluster of hits occurring early in time, a large time gap, and then another cluster of hits occurring later.  To remove such cases, the largest time gap between consecutive hits was required to be small, defined as follows.  The quantities $f_{xt}$ and $f_{yt}$ were calculated for each track.  For each view, $f$ is the ratio of the largest time gap between hits in the track to the total extent of the track in time.  A high-quality track has a value of $f$ close to zero, whereas a track built from two unrelated cosmic rays will have a value of $f$ close to unity.  The maximum of $f_{xt}$ and $f_{yt}$ is called $f_{\mathrm{max}}$.

\begin{figure}
\begin{center}
\includegraphics[width=\figurescale]{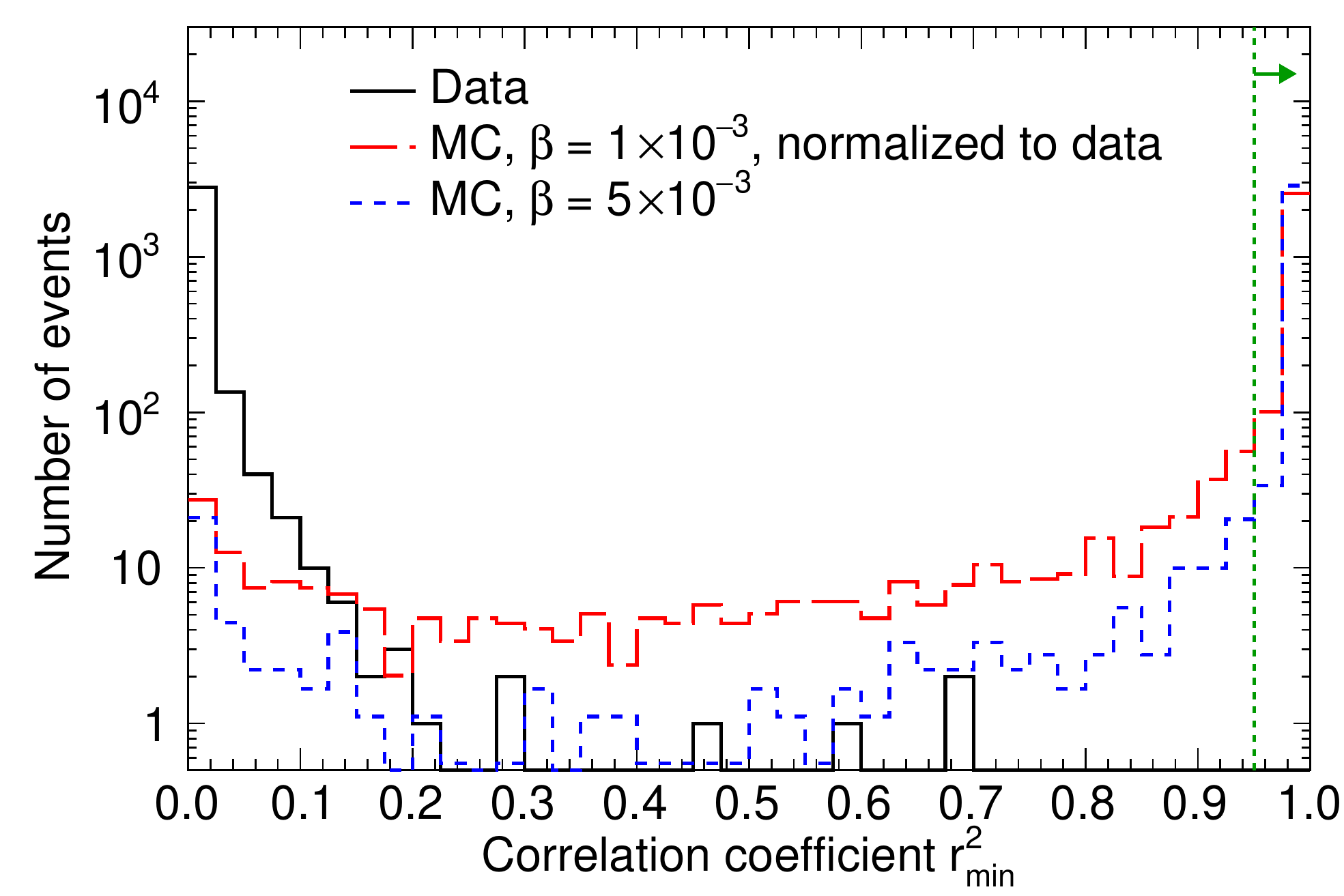}
\includegraphics[width=\figurescale]{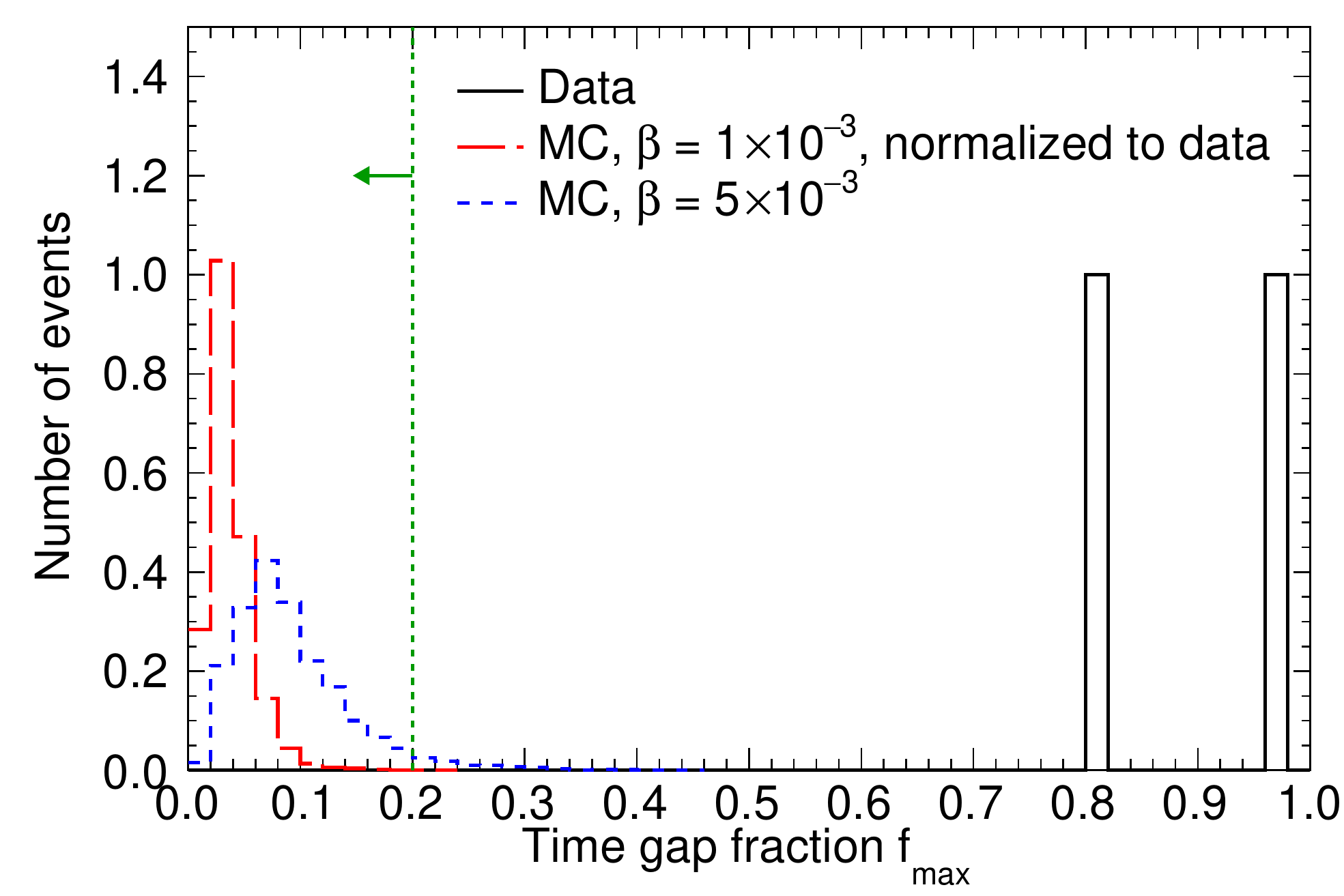}
\caption{Correlation coefficient $r^2_{\mathrm{min}}$ (top) and the time gap fraction $f_{\mathrm{max}}$ (bottom) of monopole candidates.  In each plot, events are shown only if they pass every selection except the one shown in that plot (displayed as a dashed line with arrow), i.e. for the top (bottom) plot, selections 1--4 and 6 (1--5) were applied.  Each Monte Carlo simulation is normalized to the data for display purposes.}
\label{fig:r2_gap}
\end{center}
\end{figure}

The selection criteria were chosen using 1\% of the data collected, under the assumption that such samples would contain no monopoles, since it is already known that monopoles, if they exist, are rare.  The remaining criteria for selecting an event as a monopole were:

\begin{enumerate}
\item[4.] $\beta < 10^{-2}$;
\item[5.]
$r^2_{\mathrm{min}} \ge 0.95$; and
\item[6.] 
$f_{\mathrm{max}} \le 0.2$.
\end{enumerate}
Figure~\ref{fig:r2_gap} shows the strength of $r_\mathrm{min}^2$ and $f_\mathrm{max}$ in separating signal from background.  The cutoff values for these variables were not highly optimized, but rather chosen by hand to clearly lie far from the background while retaining most of the signal.

As shown, the $f_\mathrm{max}$ cut is very efficient for $\beta = 10^{-3}$; this high efficiency holds for slower speeds as well.  At larger $\beta$, there is some loss of efficiency because it becomes possible for larger contiguous sections of monopole tracks to overlap in time with speed-of-light background tracks.  Since these sections are removed, they appear as time gaps.  In contrast, the $r^2_{min}$ cut becomes more efficient at high $\beta$ because the detector crossing time is lower and it is more likely that a simulated monopole track is entirely uncontaminated by non-monopole background hits.

Finally, we planned to visually examine any event passing all selections to determine if it appeared to be an unanticipated background.

\section{Results}\label{sec:results}

The data set recorded from 5 June 2015 through 12 October 2015 was used for this search.  During this period of time the FD was set to a consistent APD gain setting which was changed for later data taking. In these 129~days, the detector provided good data for $8.21\times10^6$\,s.  The data-driven trigger system was 99.9\% efficient during this time period, where the small inefficiency was caused by the available CPU time for the trigger to process incoming data being exhausted.  The corrected livetime is therefore $8.20 \times 10^{6}$\,s (94.9~days).

\begin{figure}
\begin{center}
\includegraphics[width=\figurescale]{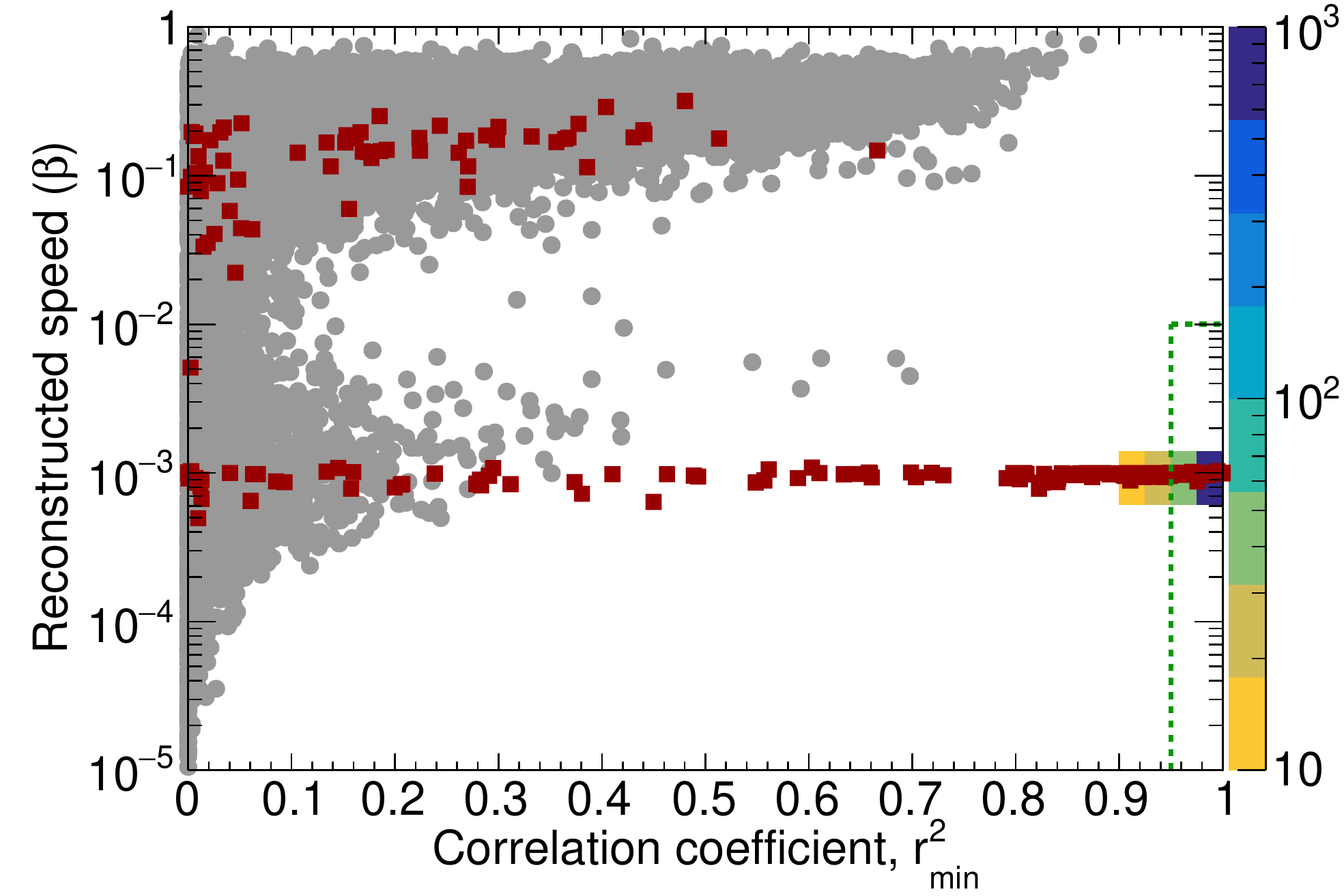}

\includegraphics[width=\figurescale]{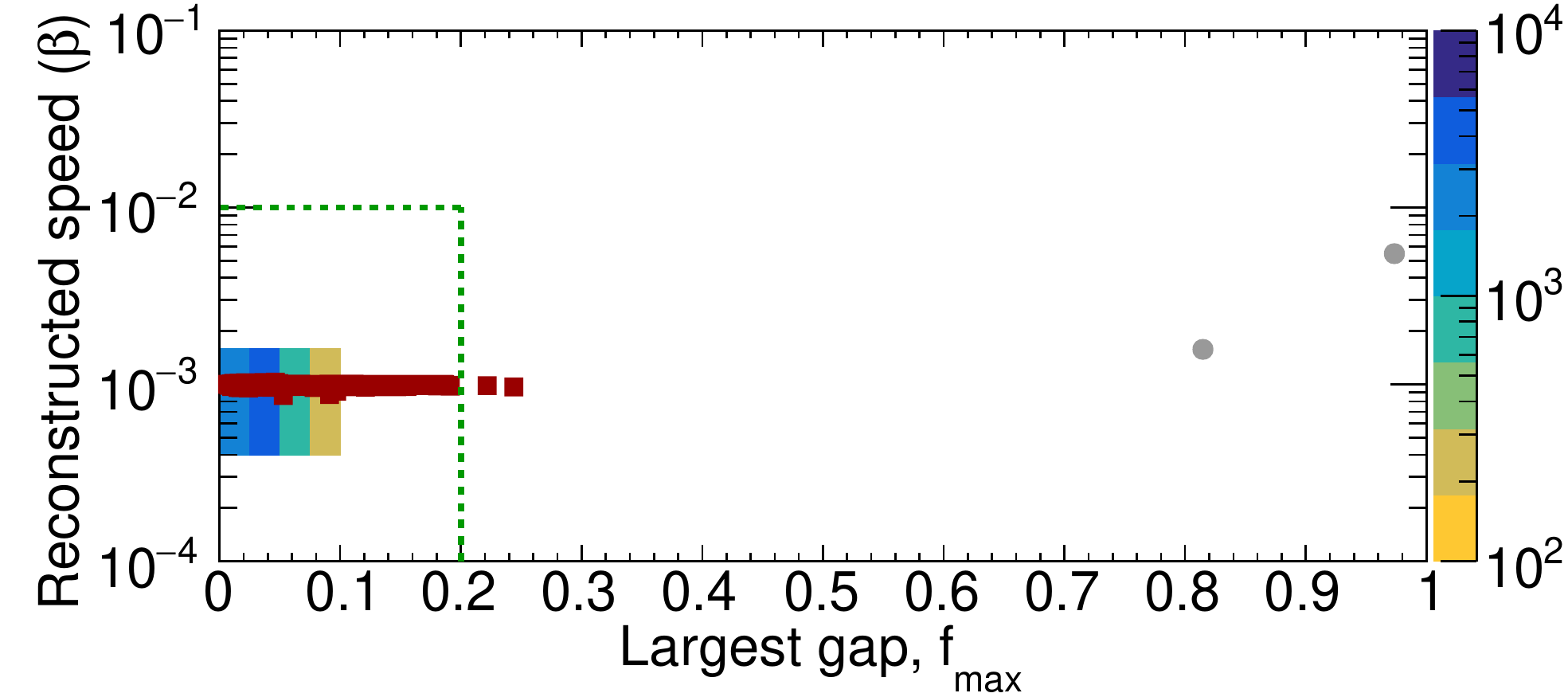}

\caption{Top: Reconstructed monopole speed vs.\ $r^2_\mathrm{min}$ for events passing selections 1--3 and 6. Data events are shown as grey circles, and simulation for $\beta = 10^{-3}$ both as red squares and as a heat map.  Events must be in the dashed box in the lower right to be selected.  Bottom: Same for $f_\mathrm{max}$ and events passing selections 1--3 and 5.  Events must be in the lower left to be selected.}
\label{fig:scatter}
\end{center}
\end{figure}

The data set contained 10\,447\,881 events, none of which fell into the signal region.  Figure~\ref{fig:scatter} shows the distribution of the full sample in speed vs. $r_\mathrm{min}^2$ and speed vs. $f_\mathrm{max}$.  All data events are far from the signal region.  The two large clusters of background events in the data around $\beta = 10^{-3}$ and $\beta = 0.5$ were caused by electronics effects and speed-of-light muons not removed in the first reconstruction step, respectively.  The former, which gives candidate tracks reconstructed within the target $\beta$ range but with very poor $r^2_{min}$ values, was confined to modules with recent large energy depositions.  It is caused by a voltage overshoot in the electronics which retriggers most or all channels on a board at a fixed time after the energy deposition.  This tends to form rectangular clusters of spurious hits rather than lines, which explains the $r^2_{min}$ values typically around 0.1 and never higher than $\sim$0.4.  Although in this analysis we did not filter out such spurious hits, they could be identified in a future analysis by their proximity in time and space to the instigating cosmic shower.

The most signal-like events were reconstructed with $\beta$ between $10^{-3}$ and $10^{-2}$ and $r^2_\mathrm{min}$ around 0.65.  Upon visual inspection, these events proved to be caused either by two speed-of-light tracks in the same location at slightly different times, or by tracks formed out of fragments of high energy showers.  In neither case would other events with the same characteristics easily be able to satisfy the requirement of $r^2 >0.95$.

In the absence of any candidates in the signal region, the 90\% C.L. flux upper limit is $\Phi_{90\%} = 2.3/L$, where $L = \Omega \epsilon A t$ is the integrated product of acceptance and livetime, $\Omega$ is the solid angle coverage, $\epsilon$ is the efficiency, $A$ is the projected surface area of the FD visible to the monopole, and $t$ is the integrated livetime.  Each quantity is detailed below.

\begin{figure}
\centerline{\includegraphics[width=\figurescale]{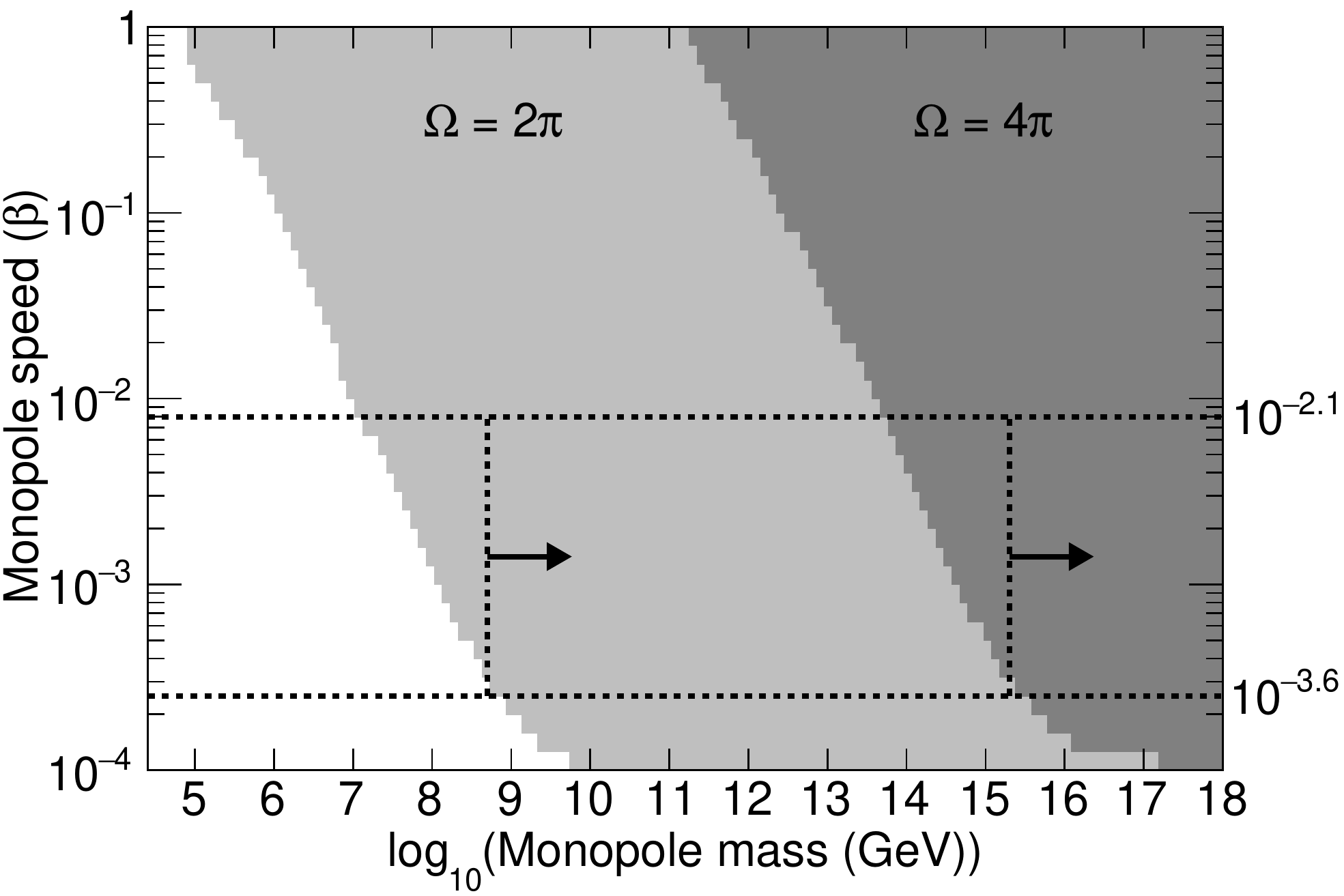}}
\caption{Solid angle coverage as a function of monopole speed and mass.  Horizontal dashed lines indicate the boundaries of the search sensitivity in speed.  Vertical dashed lines indicate the reference masses used to generate limits shown in Fig.~\ref{fig:limit_plot} and Table~\ref{tab:limits}.}
\label{fig:coverage}
\end{figure}

Limits are reported for the two major coverage scenarios: half coverage where $\Omega = 2 \pi$, and full coverage where $\Omega = 4 \pi$. The coverage depends on the kinetic energy of the monopole, which is calculated from the monopole's speed and mass.  If the monopole's energy were sufficient, it could traverse the entire planet.  In this case, the FD has $4\pi$ coverage.  The half coverage regime occurs if the monopole had enough energy to make it through the atmosphere from above, but not enough to reach the FD from below.  Figure~\ref{fig:coverage} shows the solid angle coverage as a function of monopole speed and mass.

We calculate the detector's projected area, $A$, for each simulated monopole's trajectory.  The reconstruction efficiency also depends on this trajectory, so for each monopole speed, the product was determined event by event, $\epsilon A \equiv \langle \epsilon_i A_i \rangle$.

The overall efficiency, considering both trigger efficiency and analysis selection, was 53\% (see Fig.~\ref{fig:thetax}), with most of the loss at the analysis stage arising from non-monopole hits being associated with monopole tracks and spoiling their $r^2$ values.  As shown in Fig.~\ref{fig:scatter} (top), there are two distinct topologies.  Either a small number of non-monopole hits are attached to a simulated monopole track, in which case the speed is still reconstructed near $\beta=10^{-3}$, or a small number of simulated monopole hits are associated with a speed-of-light particle, in which case the $r^2$ and speed are both far from the signal region.

Table~\ref{tab:limits} shows flux limits as a function of $\beta$.  Limits are shown for two mass cutoffs, $5\times10^8$\,GeV and $2\times10^{15}$\,GeV.  The former is the smallest mass for a monopole that would reach the detector through the atmosphere and detector overburden alone at the lower limit of the range of speeds considered, $\beta = 10^{-3.6}$. The latter is the smallest mass that would reach the detector through the Earth at $\beta = 10^{-3.6}$.  These cutoffs are shown as vertical dashed lines in Fig.~\ref{fig:coverage}.  At a given speed, the flux limits are valid for the respective shaded regions of this figure, which include somewhat lower masses than the two cutoff values for all $\beta > 10^{-3.6}$.  For instance, the $\Omega = 2\pi$ limit for $\beta = 10^{-2.1}$ is valid for masses $> 2\times 10^7$\,GeV.

\begin{figure}
\centerline{\includegraphics[width=\figurescale]{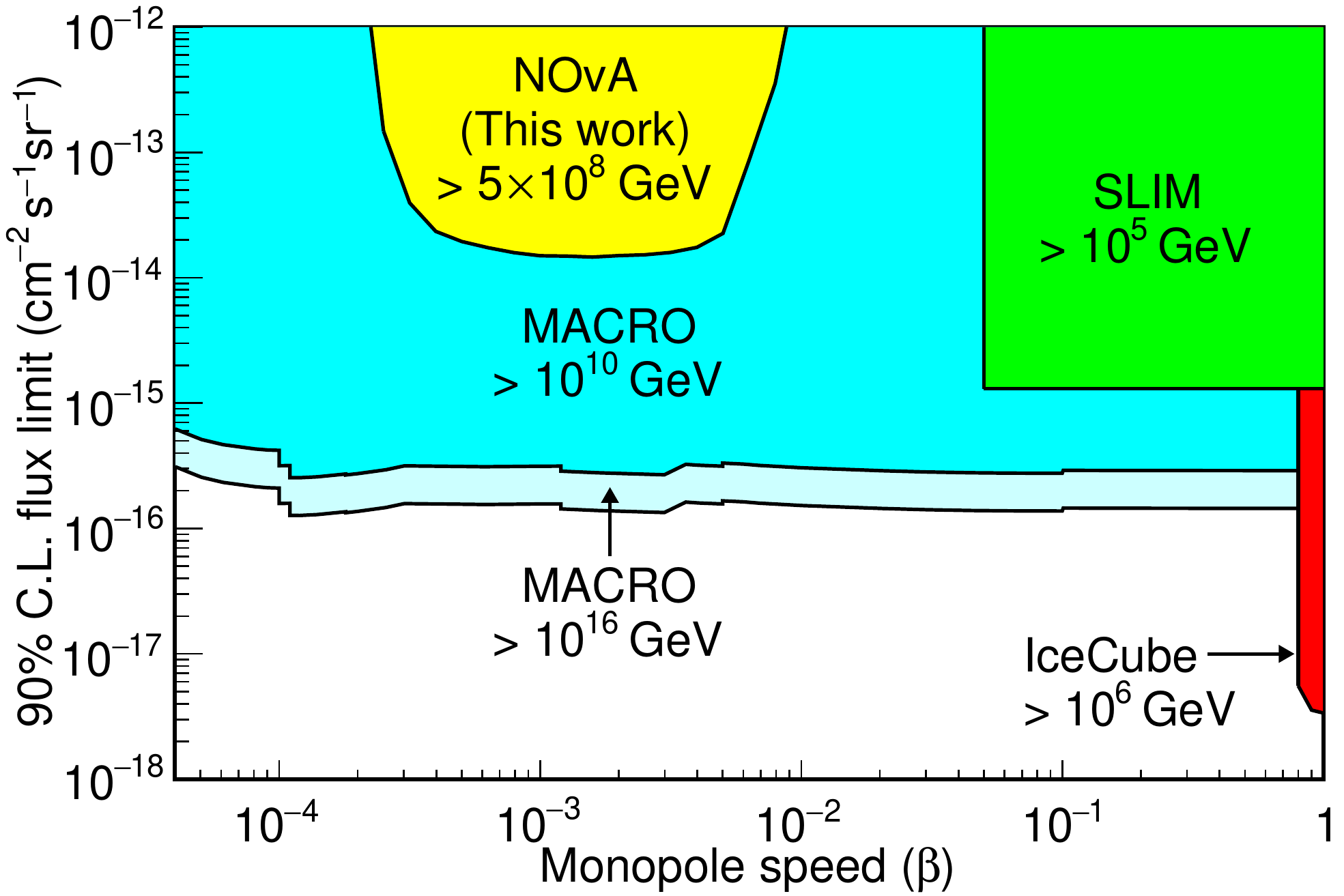}}
\caption{Upper limits on the magnetic monopole flux. Results are shown of experiments~\cite{macro,slim,icecubefast} that do not assume proton decay catalysis~\cite{icecubeslow} nor ultrarelativistic monopoles~\cite{rice}.  In each region of speed-flux space, the experiment with the best mass reach is shown.  NOvA sets the only limits around $\beta = 10^{-3}$ for monopoles lighter than $10^{10}$\,GeV.}
\label{fig:limit_plot}
\end{figure}

Sensitivity falls off at low $\beta$ as monopole energy deposition drops below the analysis threshold, given the assumption of a monopole with a single Dirac unit $g$ of charge.  This assumption is shared by the previous experiments whose limits are displayed in Fig.~\ref{fig:limit_plot}.  At high $\beta$, the sensitivity was limited by the trigger design. 

\begin{table}
{
\setlength{\tabcolsep}{9pt}
\caption{90\% C.L. upper limits on the magnetic monopole flux, in units of $10^{-15}\,\mathrm{cm^{-2}s^{-1}sr^{-1}}$.}\begin{center}
\begin{tabular}{c c c}
\hline
\hline 
$\beta$ & $m > 5\times10^8$\,GeV & $m > 2\times10^{15}$\,GeV \\
\hline
  $10^{-3.6}$ & 150 & \phantom{0}74\phantom{.0} \\
  $10^{-3.5}$ & \phantom{0}40 & \phantom{0}20\phantom{.0} \\
  $10^{-3.4}$ & \phantom{0}23 & \phantom{0}12\phantom{.0} \\
  $10^{-3.3}$ & \phantom{0}19 & \phantom{00}9.7 \\
  $10^{-3.2}$ & \phantom{0}17 & \phantom{00}8.7 \\
  $10^{-3.1}$ & \phantom{0}16 & \phantom{00}7.9 \\
  $10^{-3.0}$ & \phantom{0}15 & \phantom{00}7.5 \\
  $10^{-2.9}$ & \phantom{0}15 & \phantom{00}7.4 \\
  $10^{-2.8}$ & \phantom{0}15 & \phantom{00}7.3 \\
  $10^{-2.7}$ & \phantom{0}15 & \phantom{00}7.5 \\
  $10^{-2.6}$ & \phantom{0}15 & \phantom{00}7.6 \\
  $10^{-2.5}$ & \phantom{0}16 & \phantom{00}8.0 \\
  $10^{-2.4}$ & \phantom{0}17 & \phantom{00}8.7 \\
  $10^{-2.3}$ & \phantom{0}23 & \phantom{0}11\phantom{.0} \\
  $10^{-2.2}$ & \phantom{0}86 & \phantom{0}43\phantom{.0} \\
  $10^{-2.1}$ & 360 & 180\phantom{.0} \\
\hline\hline
\end{tabular}
\end{center}
\label{tab:limits}
}
\end{table}

These limits are conservative since they use the lower bound on hit efficiency as described in Section~\ref{sec:simulation}.  A higher efficiency would extend the quoted limits slightly towards lower masses. Other detector-based systematic uncertainties were negligible.  We considered uncertainties in livetime and solid angle and found that each was well under 1\%.  However, substantial theoretical uncertainty in slow monopole $dE/dx$ feeds into an uncertainty on translating non-detection into a flux limit.  As stated above, these limits use the nominal $dE/dx$ of Ref.~\cite{Ahlen:1982mx}.  Higher or lower energy depositions would modify the limits in the same way as changes in detector efficiency.

\section{Conclusion}

By virtue of being a large segmented detector on the Earth's surface, the NOvA FD is uniquely sensitive to slow monopoles in the mass range below $10^{10}$\,GeV which would not have reached previous detectors such as MACRO.  We have constrained the flux of this population of monopoles in a large region of speed-mass space which has previously been unconstrained, setting an upper limit on the flux of $2\times 10^{-14}\,\mathrm{cm^{-2}s^{-1}sr^{-1}}$ at 90\% C.L. for $6\times 10^{-4} < \beta < 5\times 10^{-3}$ and mass greater than $5\times 10^{8}$\,GeV.

The results shown here represent less than 10\% of the data the NOvA Far Detector has collected to date.  The data sets collected beginning in October 2015 have a higher APD gain setting, which allows collection of fainter signals and thus an improved mass reach for magnetic monopoles with $\beta < 0.01$.  NOvA also has the capability of searching for monopoles with $\beta > 0.01$~\cite{Wang:2015ery}.

This document was prepared by the NOvA collaboration using the resources of the Fermi National Accelerator Laboratory (Fermilab), a U.S. Department of Energy, Office of Science, HEP User Facility. Fermilab is managed by Fermi Research Alliance, LLC (FRA), acting under Contract No. DE-AC02-07CH11359. This work was supported by the U.S. Department of Energy; the U.S. National Science Foundation; the Department of Science and Technology, India; the European Research Council; the MSMT CR, GA UK, Czech Republic; the RAS, RFBR, RMES, RSF, and BASIS Foundation, Russia; CNPq and FAPEG, Brazil; STFC, and the Royal Society, United Kingdom; and the state and University of Minnesota.  We are grateful for the contributions of the staffs of the University of Minnesota at the Ash River Laboratory and of Fermilab.

\bibliographystyle{apsrev4-1}
\bibliography{monopole}
 
\end{document}